# DIAGNOSTICS UPGRADES FOR INVESTIGATIONS OF HOM EFFECTS IN TESLA-TYPE SCRF CAVITIES*

A.H. Lumpkin#, D. Edstrom Jr., J. Ruan, R. Thurman-Keup Y. Shin, P. Prieto, N. Eddy
Fermi National Accelerator Laboratory, Batavia, IL 60510 USA
B.E. Carlsten**, Los Alamos National Laboratory, Los Alamos, NM 87545 USA

*Abstract*

We describe the upgrades to diagnostic capabilities on the Fermilab Accelerator Science and Technology (FAST) electron linear accelerator that will allow investigations of the effects of high-order modes (HOMs) in SCRF cavities on macropulse-average beam quality. We examine the dipole modes in the first pass-band generally observed in the 1.6-1.9 GHz regime for TESLA-type SCRF cavities due to uniform transverse beam offsets of the electron beam. Such cavities are the basis of the accelerators such as the European XFEL and the proposed MaRIE XFEL facility. Preliminary HOM detector data, prototype BPM test data, and first framing camera OTR data with ~20-micron spatial resolution at 250 pC per bunch will be presented.

## INTRODUCTION

There are current Free-Electron Laser (FEL) initiatives that will be enabled by the use of the TESLA-type SCRF cavities in Europe (the European XFEL) and proposed in the USA (the MaRIE facility). One of the challenges is the control of the high-order modes (HOMs) that develop in these cavities due to transverse beam offsets . Diagnostic capabilities are being upgraded on the Fermilab Accelerator Science and Technology (FAST) facility [1] that will allow investigations of the effects of high-order modes (HOMs) in SCRF cavities on macropulse-average beam quality. We focus on the dipole modes in the first pass-band generally observed in the 1.6-1.9 GHz regime in TESLA-type SCRF cavities due to beam offsets. Such cavities are the basis of the accelerators for the European XFEL and the proposed MARIE XFEL facility. Raw HOM data indicate that the mode amplitudes oscillate for ~10 μs after the micropulse enters the cavity. With a 3-MHz pulse train, we expect transverse centroid shifts will then occur during the macropulse resulting in a blurring of the beam-size image averaged over the macropulse.

To evaluate these effects, upstream corrector magnets were tuned to steer the beam off axis upon entering the first of the two cavities. From there, several parameters were tracked, including the two HOM detector signal strengths for each cavity, the average transverse beam positions from rf beam position monitors (BPMs), and the average beam size using intercepting screens and imaging. Preliminary efforts demonstrated reduction of the HOM signals by beam steering. Our initial data from an optical transition radiation (OTR) source indicated a framing camera mode can provide ~20-micron spatial resolution at ~250 pC per bunch while the prototype rf BPM required higher charge to approach this resolution for single-bunch beam position. The preliminary HOM detector data, prototype BPM test data, and framing camera data will be presented later in this paper.

## EXPERIMENTAL ASPECTS

### *The FAST Linac*

The FAST linac is based on the L-band rf photocathode (PC) gun which injects beam into two superconducting rf (SCRF) capture cavities denoted CC1 and CC2, followed by transport to a low energy electron spectrometer. A $Cs_2Te$ photocathode is irradiated by the UV component of the drive laser system described elsewhere [2]. The basic diagnostics for the HOMs studies include the rf BPMs located before, between, and after the two cavities as shown in Fig. 1. These are supplemented by the imaging screens at X107, X108, X121, and X124. The HOM couplers are located at the upstream and downstream ends of each SCRF cavity, and these signals are processed by the HOM detector circuits with the output provided online though ACNET, the Fermilab accelerator controls network. The upgrades will include optimizing the HOM detectors' bandpass filters, reducing the 1.3 GHz fundamental with a notch filter, converting the rf BPMs electronics to bunch-by-bunch capability with reduced noise, and using the C5680 streak camera in a rarely-used framing mode for bunch-by-bunch spatial information.

### *The Streak Camera System*

In an initial framing camera study [3] we observed the green component remaining from UV-conversion at 3 and 9 MHz micropulse frequencies with the laser lab streak camera. We have also recently applied the principle to optical transition radiation (OTR) from an Al-coated Si substrate with subsequent transport to a beamline streak camera that views OTR from the X121 screen location. Commissioning of the streak camera system was facilitated through a suite of controls centered around ACNET. This suite includes operational drivers to control and monitor the streak camera as well as Synoptic displays to facilitate interface with the driver. Images are captured from the streak camera using the readout cameras, Prosilica 1.3 Mpixel cameras with 2/3" format, and may be analyzed both online with a Java-based ImageTool and an offline MATLAB-based ImageTool processing program [4,5]. Bunch-length measurements using these techniques have been reported previously from the A0 Facility [6] and FAST first system streak camera commissioning at 20 MeV [7].

*This manuscript has been authored by Fermi Research Alliance, LLC under Contract No. DE-AC02-07CH11359 with the U.S. Department of Energy, Office of Science, Office of High Energy Physics.
**Work at LANL supported by US Department of Energy through the LANL/LDRD Program.
#lumpkin@fnal.gov

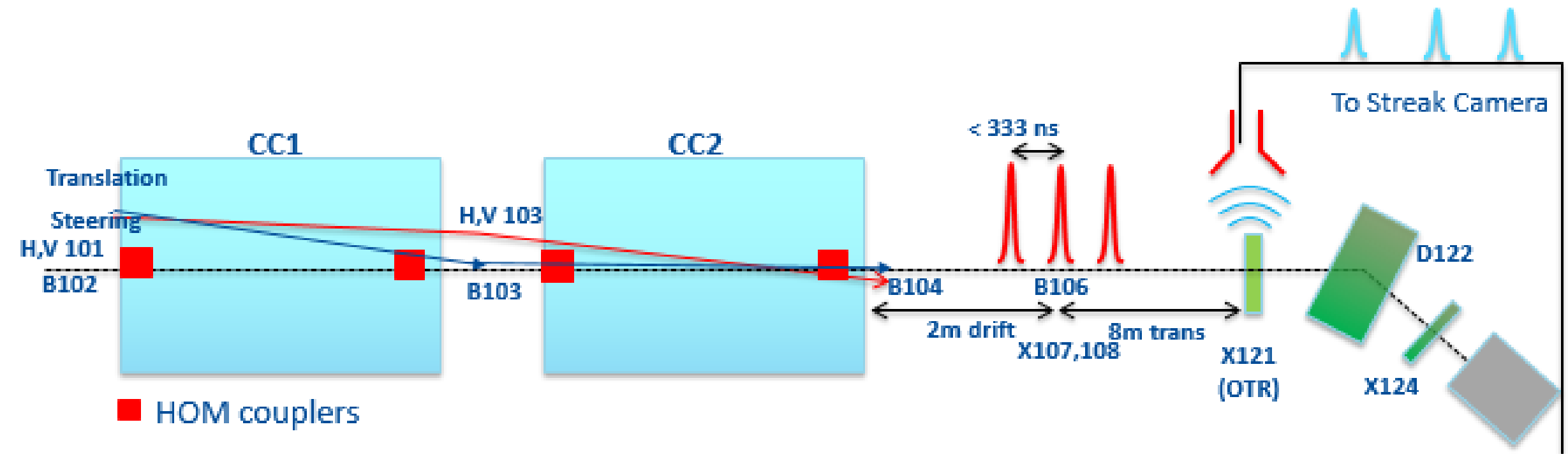

Figure 1: Schematic of the FAST beamline layout showing the capture cavities, correctors, rf BPMs, HOM couplers, X121 OTR screen, spectrometer, and path to streak camera.

The streak camera stations each include a Hamamatsu C5680 mainframe with S20 PC streak tube and can accommodate vertical sweep plugin units and either a horizontal sweep unit or a blanking unit. The UV-visible input optics allow the assessment of the 263-nm component as well as the amplified green component or IR components converted to green by a doubling crystal. The framing mode studies required replacing the M5675 synchroscan unit with the M5677 slow vertical sweep unit (5-ns to 1-ms ranges) [8]. The M5679 dual axis plugin which provides a horizontal sweep with selectable ranges had already been installed for the previous dual sweep synchroscan tests. A second set of deflection plates in the streak tube provides the orthogonal deflection for the slower time axis in the 100-ns to 10-ms regime. These plates are driven by the dual-axis sweep unit which was also commissioned during previous studies.

## EXPERIMENTAL RESULTS

### *Initial HOM Detector Data: 3 MHz*

The basic tests involved the HOM couplers installed in the upstream and downstream ends of the capture cavities, CC1 and CC2. These are designed to damp the HOMs, but in addition their signals can be filtered, processed, amplified, and digitized. In our initial tests, we used a bandpass filter at 1.7-1.8 GHz to select some of the dipole modes that are seen when the beam transits the cavities off axis. We used the settings, as found, for steering through the cavities. Sample waveforms are shown in Fig. 2 for the steering normally used during the Summer of 2016. However, we show that the CC1 detector signals can be reduced and minimized by a factor of 5 by steering with 1.15 A in corrector H101 as seen in Fig. 3. We subsequently steered H and V 103 to minimize the HOMs in CC2 as well. When these new corrector settings were then used a few days later, the observed x,y emittances were reduced by 20-30%. In the upgrade, we are using a notch filter at 1.3 GHz to reduce it by 10,000, wider bandpass filters at 1.6-1.9 GHz for the dipole modes, and a shortpass filter at 2.2 GHz to reduce any signal from the monopole modes in the cavities.

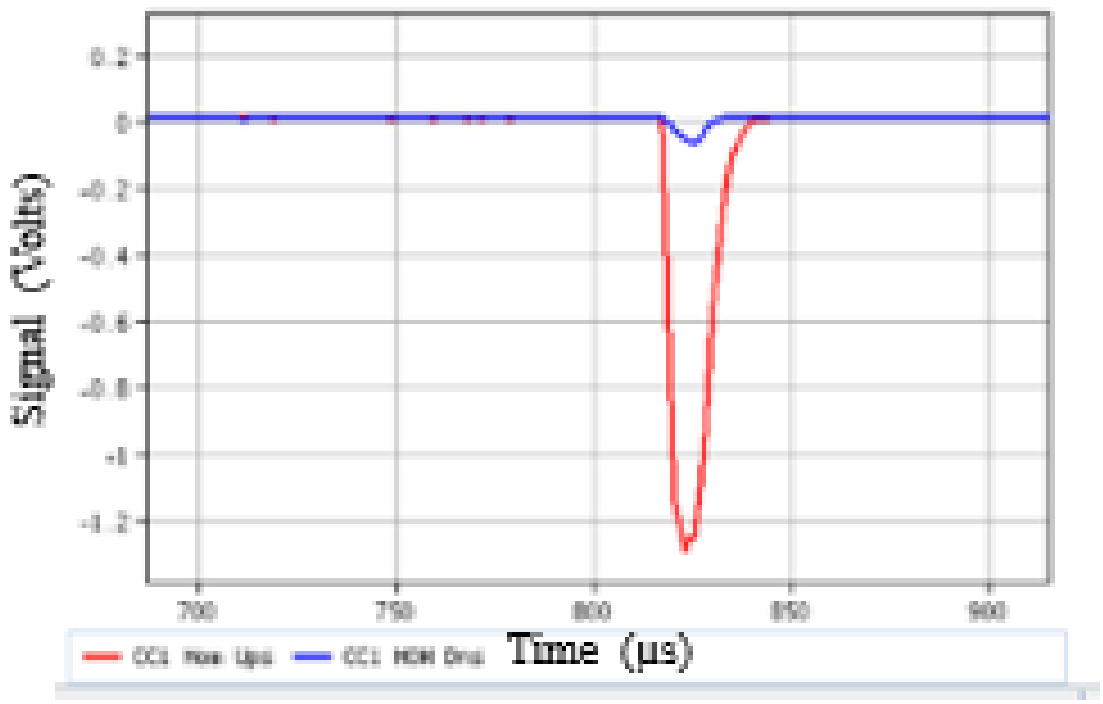

Figure 2: Digitized signal envelopes from the CC1 upstream (red) and downstream (blue) detectors with 750 pC/bunch and a 25-µs-long pulse train.

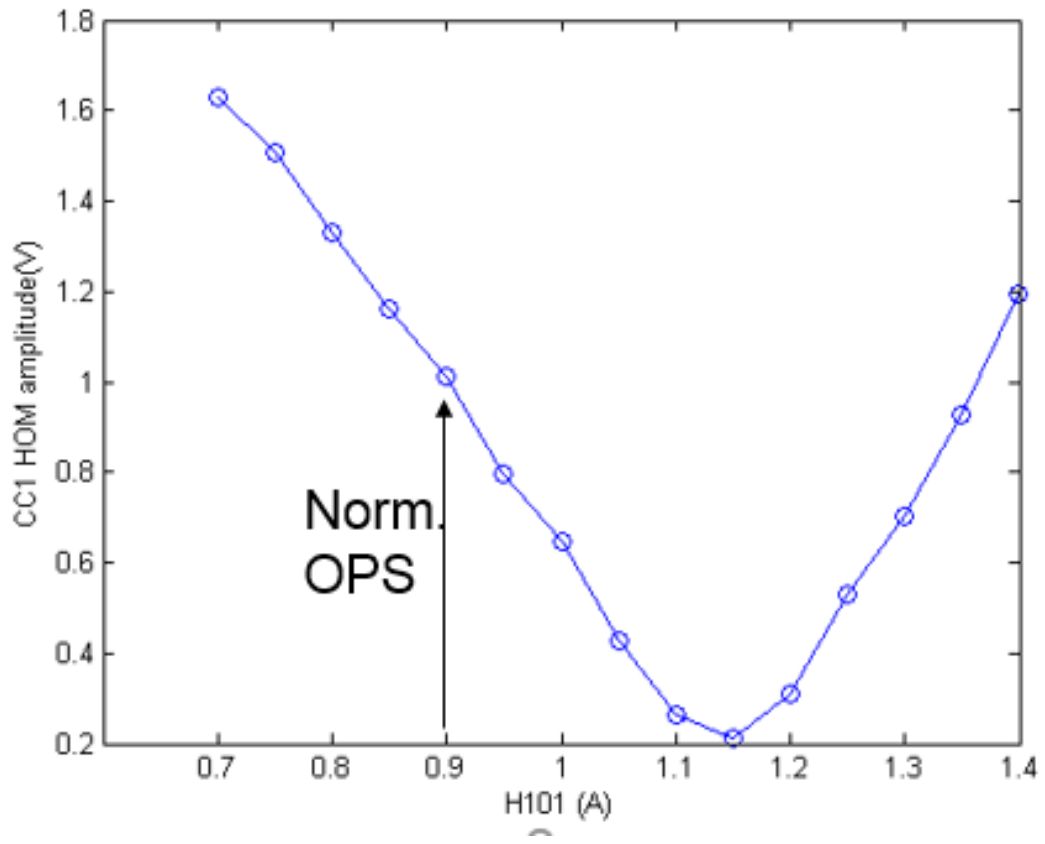

Figure 3: CC1 HOM detector 1 peak signal variation with H101 corrector settings.

Another preliminary trajectory study was performed by adjusting the corrector H101 located ~0.5 m upstream of the first capture cavity CC1. We tracked the average rf BPM readings in B102, B103, B104, and B106 as a function of the H101 corrector current settings. The data in Fig. 4 indicate that the normal setting of 0.9 A corresponded to a 1.4-mm offset at B102 as beam enters CC1. The B103 readings are somewhat insensitive to these entrance values due to the focusing effects by cavity fields in CC1. More significant mm-scale offsets are seen after CC2, and one would expect stronger HOMs in CC2.

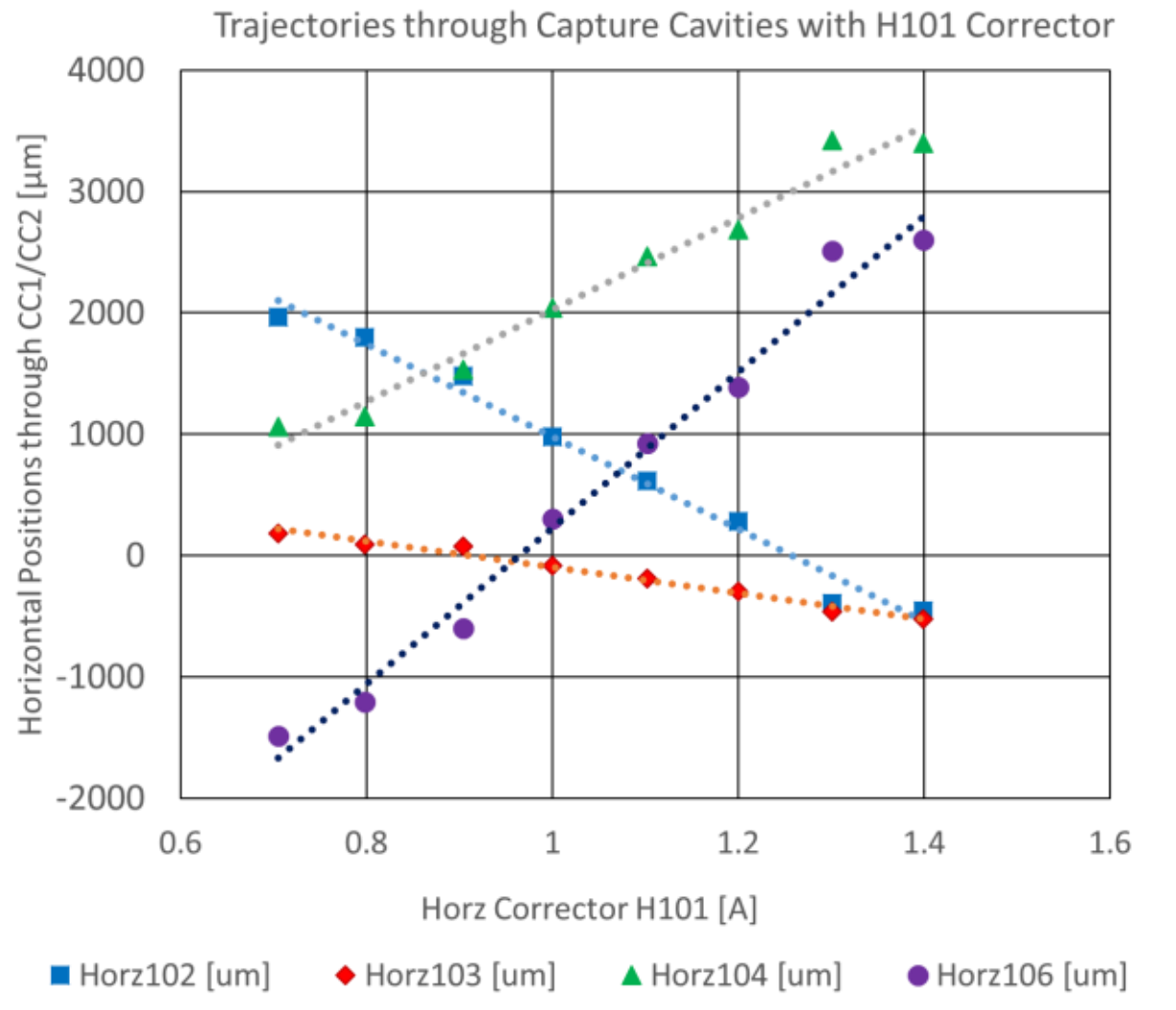

Figure 4: Sample of tracking of the average beam centroids through CC1 and CC2.

## *Bunch by Bunch rf BPM*

The upgrades in the prototype rf BPM electronics board included noise reduction which enabled an improved spatial resolution for a given charge. At 2 nC per micropulse the rms noise was found to be 25 µm in x and 15 µm in y in B101 in the August 29, 2016 test with 4.5 MeV beam from the gun. The firmware to allow the tracking of the beam positions bunch by bunch was also implemented. An example of the tracking of the 50 micropulses in a macropulse train is shown in Fig. 5 before the low-noise revision was employed. Both noise-reduction and bunch-by-bunch capabilities are needed for the proposed HOM long-range wakefield studies in which the mode oscillations are anticipated to kick different micropulses varying amounts depending on the amplitude.

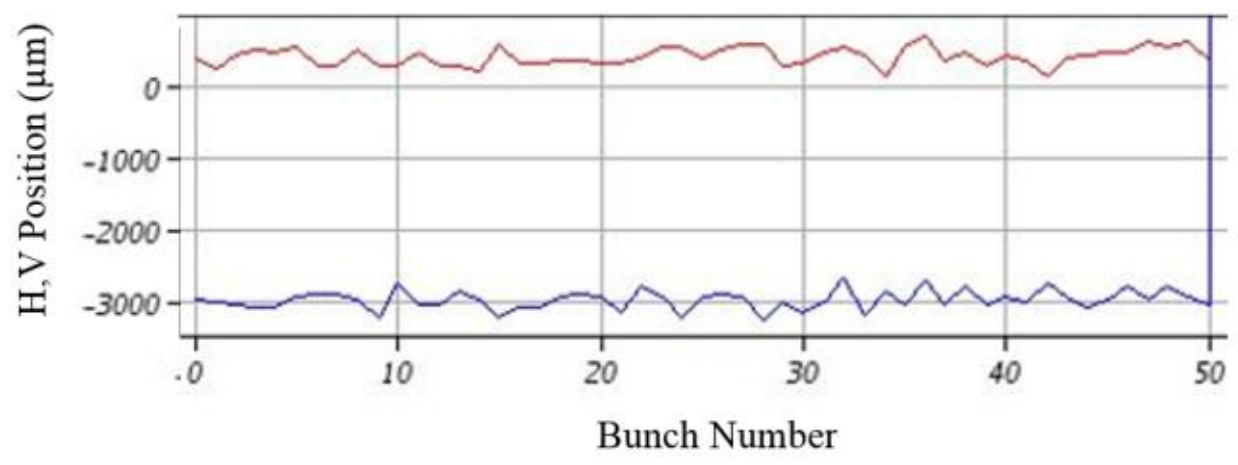

Figure 5: Sample of the tracking of the beam centroids bunch by bunch for 50 micropulses.

## *Initial Green Component studies: 3 MHz*

The basic principles are illustrated with the Laser Lab Streak Camera. In Fig. 6, a 10-µs vertical sweep shows each of the 20 micropulses from the incident laser macropulse. The vertical projected profiles for the region of interest (ROI) would show all 20 micropulses at a 3-MHz rate. Such an image could track bunch-by-bunch centroid motion as well, particularly in the horizontal plane as it is perpendicular to the sweep direction. To obtain more pulse separation vertically, one can reduce the coverage to 1 µs with a faster deflection and add the horizontal sweep to cover more micropulses.

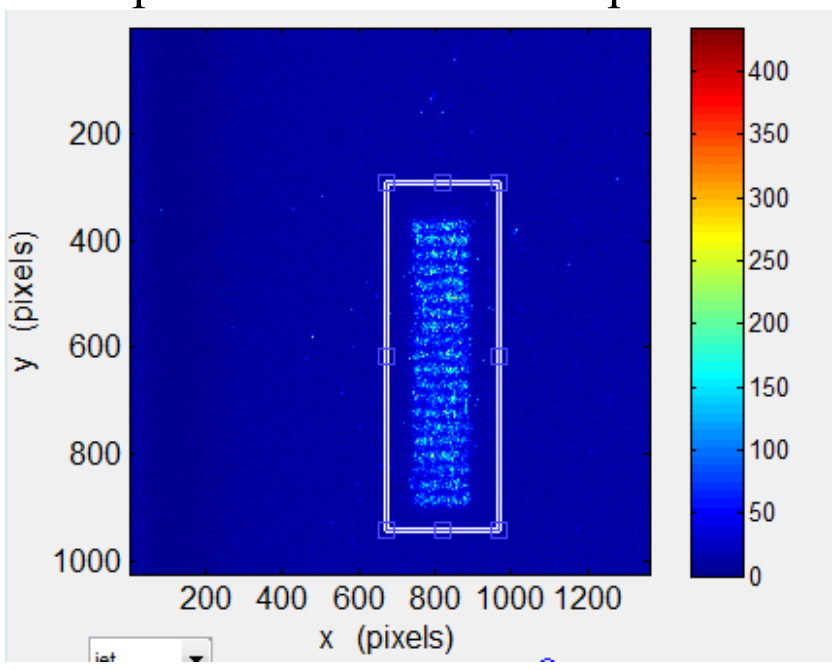

Figure 6: Vertical slow sweep image of 20 micropulses with 10-µs sweep range. Green at 3 MHz.

## *Linac Beamline Streak Camera with OTR*

The beamline streak camera is installed in an optical enclosure outside of the beamline enclosure with transport of OTR from the instrumentation cross at beamline location 121 (X121) as described [9]. The all-mirror transport minimizes the chromatic temporal dispersion effects for bunch length measurements. The same transport can be used for the framing-mode tests. In this case, we only used the horizontal sweep to separate the micropulses at 3 MHz. The input image was apertured by the entrance slits to provide $\sigma_x$=135 µm and $\sigma_y$=25 µm effective sizes for the demonstration shown in Fig. 7. The spatial resolution for the system is ~10 to 15 µm with an effective calibration factor of 6.6 µm/pixel. The OTR signal from the initial micropulse charge of ~300 pC was reduced by the apertures for the camera images. This proof of principle of the framing technique using OTR can be applied to HOM dipole mode effects within the e-beam macropulse and turn-by-turn OSR effects in IOTA.

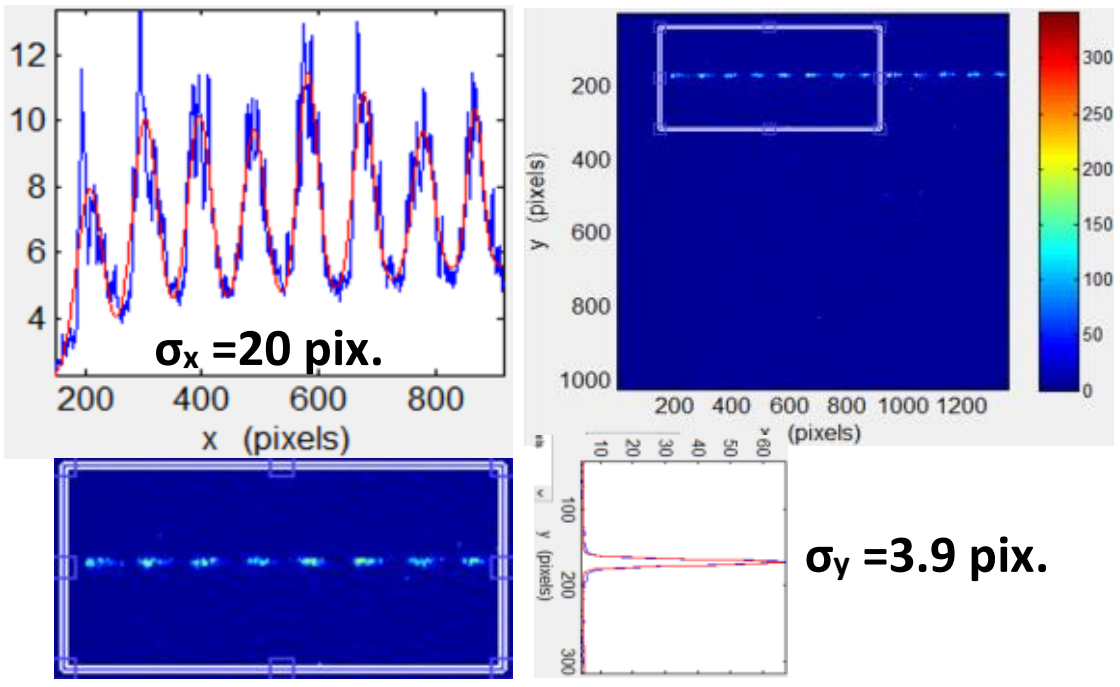

Figure 7: Semi-framing-camera mode with X121 OTR from the electron beam pulse train at 3 MHz.

## SUMMARY

In summary, a series of preliminary observations were made using the HOM detectors, the rf BPMs, and the Laser Lab streak camera configured in framing mode to track beam size and position at 3 MHz. The framing mode was also applied to OTR from X121, demonstrating viability for HOM-effect measurements upon individual e-beam micropulses. Optical synchrotron radiation sources in IOTA [10] are calculated to be brighter turn by turn than the OTR from a single micropulse in the electron beamline so we have established the proof of principle for that application as well.

## ACKNOWLEDGMENTS

The authors acknowledge the support of A. Valishev, D. Broemmelsiek, and R. Dixon, all at Fermilab. They also acknowledge support from the rest of the FAST/IOTA group and Electrical, Mechanical, Controls, and Cryogenic departments.